\documentclass[aps,prd,twocolumn,showpacs,eqsecnum,amsmath,nofootinbib]{revtex4}
\usepackage[T1]{fontenc}
\usepackage[utf8]{inputenc}
\usepackage{lmodern}
\usepackage[english]{babel}
\usepackage{xcolor}
\usepackage{graphicx,amssymb,amsthm,amsmath}
\usepackage{hyperref}
\DeclareMathOperator{\arccoth}{arcCoth} 
\def \d {{\rm d}}

\begin{document}

\title{Scalar Hairy Black Holes in the presence of Nonlinear Electrodynamics}

\author{T. Tahamtan}
\email{tahamtan@utf.mff.cuni.cz}

\affiliation{Institute of Theoretical Physics, Faculty of Mathematics and 
Physics, Charles University, V~Hole\v{s}ovi\v{c}k\'ach 2, 180~00 
Prague 8, Czech Republic}
\affiliation{Institute of Physics, Silesian University in Opava, Bezru\v{c}ovo n\'{a}m. 13, CZ-74601 Opava, CZ}

\begin{abstract}
We study influence of scalar fields on Nonlinear Electrodynamics spacetimes. The investigation is carried out using both test and gravitating scalar fields. {After revisiting Einstein--Maxwell scalar field solutions we focus on analytic investigation of Nonlinear Electrodynamics scalar field spacetimes, especially Square Root Lagrangian model.} The main motivation being to understand whether certain specific signatures of scalar fields are preserved {when Nonlinear Electrodynamics as an additional source is considered}. We {show that} the regularity of horizon which is spoiled by scalar field in spherically symmetric static scalar-vacuum spacetimes {is not improved by including Nonlinear Electrodynamics or other sources satisfying certain condition. We confirm these findings using test scalar field that enables us to go beyond spherical symmetry.} 
\end{abstract}

\date{\today}

\maketitle

\section{Introduction}

Solutions to Einstein equations with scalar field source provide very useful tool for understanding relativity due to the simplicity of source. Recently, it has become evident that fields of this type really do exist (Large Hadron Collider) and play a fundamental role in the standard model of particle physics. At the same time scalar fields feature in models of dark energy and dark matter. In classical general relativity they have been used to study counter examples to black hole no-hair theorems or the Cosmic Censorship hypothesis and in many other areas as well.

Although studies of Strong Cosmic Censorship (SCC) started more then 50 years ago \cite{penros-65, penros-69, Hawking} the field is still active today and recent results \cite{Cardoso1} indicate violation of SCC in the context of highly charged black holes with a positive cosmological constant based on decay rates of perturbations outside of event horizon. These conclusions were subsequently probed by studying nonlinear effects numerically \cite{Cardoso2} and showing the possibility of extending the geometry beyond Cauchy horizon. On the other hand, in \cite{Hod} it was shown that charged massive scalar fields in dynamically formed Reissner--Nordstr\"{o}m--de Sitter black hole lead to SCC preservation.

Investigations related to no-hair theorems have both similarly long history and active present. Having naked singularity solution or irregular horizon when scalar field is present was predicted already by J. E. Chase in 1970 --- the exact statement known as a "Chase Theorem" \cite{Chase}. According to it roughly any static spherically symmetric vacuum solution minimally coupled to massless scalar field can not have a regular horizon, if there exists any horizon it would be also the locus of a curvature singularity (see \cite{Tafel} for generalization including potential for the scalar field). There is a nice review \cite{Herdeiro} on scalar no-hair theorems where they study four dimensional asymptotically flat black holes with scalar hair in various types of scalar field models coupled to gravity without additional gauge fields.

{Soon after static scalar field solution of Einstein equations was rediscovered in \cite{JNW1} (originally derived by Fisher \cite{Fisher}), R. Penney \cite{Penney1969} generalized the scalar vacuum solution to Einstein Maxwell scalar field solution albeit considering only electric field. Later, Janis et al \cite{JNW2} presented a method for generating Einstein scalar field and Einstein--Maxwell scalar field solutions out of the vacuum ones. Other branches of the Einstein--Maxwell scalar field solutions were later studied in \cite{Uhlir1973}. Subsequent works \cite{Teixeira} give a general class of static cylindrically symmetric solutions which allows combinations of electric and magnetic fields. The investigation proceeded to broaden the class of spacetimes even further and stationary axially symmetric Einstein--Maxwell scalar spacetimes were derived and a method to generate such solutions from the corresponding vacuum ones was presented \cite{Gurses1977}. Most recently \cite{Maeda2019}, a higher-dimensional generalization of static solutions for the Einstein--Maxwell system with a massless scalar field was performed using the Buchdahl and Janis--Robinson--Winicour transformations \cite{JNW1, JNW2}.}	
	 	
	  {Several authors used generating methods to obtain non-static Einstein--Maxwell scalar field solutions \cite{Roy}. In order to study Cosmic Censorship hypothesis violation, Roberts \cite{Roberts} considered dynamical solutions, especially the so-called Vaidya--Wyman spacetime. Recently \cite{Tahamtan-PRD-2015, Tahamtan2}, dynamical scalar field spacetime without symmetries was derived and its physical properties (e.g. presence of gravitational waves, Bondi mass) were investigated together with limits to previously known spherically symmetric solutions. Additionally, violations of Cosmic Censorship are analyzed therein.} 

Lately, investigations of the above mentioned crucial questions {(SCC and no-hair theorem)} in general relativity expanded to cover also Nonlinear Electrodynamics (NE). The idea of Nonlinear Electrodynamics is almost a century old and was initially developed as a solution to the problem of divergent field of a point charge (see e.g. \cite{Dirac}) also giving reasonable self-energy of charged particle. The best-known and frequently used form of NE was presented already in 1934 by Born and Infeld \cite{BornInfeld}. Excellent overview of the subject was given in a book by Pleba\'{n}ski \cite{Plebanski}. Later, other NE models were considered for both solving the point charge singularity and resolving the spacetime singularity \cite{Ayon-Beato1, Ayon-Beato2, Bronnikov, china}. 

Recently, perturbation by scalar fields with and without charge on the background of Born-Infeld-de Sitter black hole was analyzed in \cite{SCC-Born}. It was found that when nonlinearity in this specific model of Nonlinear Electrodynamics becomes very strong there is a possibility of rescuing SCC. If one checks the limit of strong nonlinearity for the Born--Infeld type model used therein the resulting form of NE is $~ \sim \sqrt{F_{\mu \nu}F^{\mu \nu}}$, with $F_{\mu \nu}$ being electromagnetic tensor. This gave us the motivation to investigate specifically the "Square Root" model of NE with scalar field while focusing mainly on the Chase theorem extension to such setting.

 {Note that the "Square Root" model has some interesting properties, that were used extensively in literature. Originally, this Lagrangian was proposed by Nielsen and Olesen \cite{Nielsen1973} in a flat spacetime to treat so-called dual string. Inspired by 't Hooft's paper \cite{Hooft} about the importance of linear potential term for models of permanent confinement,  Gaete and Guendelman \cite{Guendelman2006} showed that this confinement potential can be generated from "Square Root" term in Lagrangian of gauge theory on Minkowski background. Subsequently, this approach was generalized to curved backgrounds and also to modified theories of gravity. A nice review of recent developments on non-linear gauge theory containing "Square Root" Lagrangian can be found in \cite{Guendelman2014}. On the other hand, "Square Root" Lagrangian is a subclass of power Maxwell Lagrangian ($-F^s$) which was studied in \cite{square root} mainly on curved backgrounds and new exact solutions in four and higher dimensions were derived and compared with solutions for other models of NE.}


In this paper we investigate the existence of regular horizons in {static} spacetimes with NE by using both exact analytical calculations and perturbative approach. After describing the overall setup of the problem (part \ref{field-equations}) and briefly mentioning the solutions with single source (parts \ref{scalar-field} and \ref{NE-solution}), we briefly revisit the Einstein--Maxwell scalar field solutions \ref{EMS-solution}. Then we derive explicit solution for the "Square Root" model of NE with scalar field and investigate the relevant properties (in part \ref{Exact-sol-SC-NE}). Subsequently, we extend the analysis to more general sources than the specific NE model used before (part \ref{General-source}). Finally, we look at the scalar test field approach to understand the problem from other perspective (part \ref{Test-field}).

\section{Fields Equations}\label{field-equations}
We consider the following action, describing a scalar field minimally coupled to gravity and also NE, 
\begin{equation}\label{action}
S=\frac{1}{2}\int d^{4}x \sqrt{-g}[\mathcal{R}+\nabla_{\mu}\varphi \nabla^{\mu} \varphi+\mathcal{L}(F)]\, ,
\end{equation}
where $\mathcal{R}$ is the Ricci scalar for the metric $g_{\mu \nu}$ (we use units in which $c=\hbar=8 \pi G=1$.). The massless scalar field $ \varphi $ is supposed to be real.  $\mathcal{L}(F)$ is the Lagrangian of the Nonlinear Electromagnetic field which we assume to be an arbitrary function of the invariant $F=F_{\mu \nu}F^{\mu \nu}$ constructed from a closed Maxwell 2-form $F_{\mu \nu}$.

{Generally, the Lagrangian $\mathcal{L}$ of Nonlinear Electrodynamics is supposed to be a scalar function of the invariants $F=F_{\mu \nu}F^{\mu \nu}$ and $G=F_{\mu \nu}{}^{*}F^{\mu \nu}=\frac{1}{2}\epsilon^{\mu \nu \alpha \beta}F_{\mu \nu}F_{\alpha \beta}$ (in fact one should consider only $G^{2}$ to eliminate pseudoscalar nature of $G$). Since we are interested in static spherically symmetric and ``pure magnetic field'' (the same would apply to ``pure electric field'') solutions the second invariant $G\sim \mathbf{E\cdot B}$ vanishes identically, so we consider only Lagrangians of the form $\mathcal{L}(F)$.}

 By applying the variation with respect to the metric for the action (\ref{action}), we get Einstein equations
\begin{equation}\label{field equations}
G^{\mu}{}_{\nu}=T^{\mu}{}_{\nu},
\end{equation}
where $T^{\mu}{}_{\nu}={}^{\rm SF}T^{\mu}{}_{\nu}+{}^{\rm EM}T^{\mu}{}_{\nu}$ . The energy momentum tensor generated by the scalar field is given by
\begin{equation} \label{energy-momentum scalar-field}
{}^{\rm SF}T_{\mu \nu}=\nabla_{\mu}\varphi\, \nabla_{\nu}\varphi-\frac{1}{2}g_{\mu \nu}\,g^{\alpha\beta}\nabla_{\alpha}\varphi \nabla_{\beta}\varphi\,.
\end{equation}
and the scalar field must satisfy corresponding field equation
\begin{equation}\label{box}
\Box \varphi=0\, ,
\end{equation}
where $\Box$ is a standard d'Alembert operator for our metric ($g_{\mu \nu}$).

The electromagnetic energy momentum tensor is defined as following  
\begin{equation}\label{energy-momentum-Maxwell}
{}^{\rm EM}T^{\mu}{}_{\nu}=\frac{1}{2}\{\delta^{\mu}{}_{\nu} \mathcal{L}-(F_{\nu \lambda}F^{\mu \lambda})\mathcal{L}_F\}\,,
\end{equation}
in which $\mathcal{L}_F=\frac{d\mathcal{L}(F)}{dF}$ and superscript ${\rm EM}$ means electromagnetic.  {Specifically for the Maxwell case $\mathcal{L}=-F$ and $\mathcal{L}_F=-1$.} The electromagnetic fields are obeying the (generally) modified Maxwell (NE) field equations. The source--free nonlinear Maxwell equations are given in the following form 
\begin{eqnarray} \label{modified Maxwell}
\d \boldsymbol{F}&=&0\, , \\
\d (\mathcal{L}_F\,\boldsymbol{^{*}F})&=&0\, ,
\end{eqnarray} 
in which $\boldsymbol{^{*}F}$ is a dual of electromagnetic two-form $\boldsymbol{F}$.

We are interested in static spherically symmetric (SSS) case and for easier comparison with previous results we consider the metric in the following form
\begin{eqnarray}\label{ourmetric}
\d s^2&=&-f(r)\,\d t^2+\frac{\d r^2}{f(r)} +R(r)^2 \d {\Omega}^2\, ,
\end{eqnarray}
where $\d {\Omega}^2=\d \theta^2 +{\sin{\theta}}^2\d \phi^2$ {and subsequently we assume $t,r,\theta,\phi$ coordinate ordering.}
 The Einstein tensor with respect to this metric has the following non-zero components 
\begin{eqnarray}
G^{r}{}_{r}&=&f\,\left(\frac{R_{,r}}{R}\right)^2+\frac{R_{,r}}{R}\,f_{,r}-\frac{1}{R^2}\, , \label{Grr}\\
G^{t}{}_{t}&=&G^{r}{}_{r}+2\,f\,\frac{R_{,rr}}{R}\, , \label{Gtt}\\
{G^{\theta}}_{\theta}&=&{G^{\phi}}_{\phi}=\frac{f_{,rr}}{2}+\frac{R_{,r}}{R}f_{,r}+\frac{R_{,rr}}{R}\,f\, , \label{Gxx}
\end{eqnarray}
and the Ricci scalar with respect to our metric anzats (\ref{ourmetric}) is 

\begin{equation}\label{Ricci}
Ricci=-f_{,rr}-\frac{4}{R}\left(f\,R_{,r}\right)_{,r}-2f\left(\frac{R_{,r}}{R}\right)^2+\frac{2}{R^2}\, .
\end{equation}

Here, we briefly mentioned all the necessary field equations which we need for rest of the paper.

\subsection{Scalar Field}\label{scalar-field}

In this part, we consider scalar-vacuum case where massless scalar field is minimally coupled to gravity. This will illustrate the problem scalar field brings in a simple model. The scalar field is assumed to be a function of $r$ only ($\varphi(r)$) and the energy momentum tensor generated by the radial scalar field (\ref{energy-momentum scalar-field}) with respect to our metric ansatz becomes 
\begin{equation}\label{Scalar field}
{}^{\rm SF}T^{\mu}{}_{\nu}=\frac{f\,\varphi^2_{,r}}{2}diag\left\{-1,1,-1,-1\right\}\, .
\end{equation}
The wave equation (\ref{box}) of scalar field with respect to our metric (\ref{ourmetric}) leads to
\begin{equation}\label{box-eq}
f\,\varphi_{,r}\,R^2=const.\, ,
\end{equation}
which can be integrated to give
\begin{equation}\label{phi}
\varphi(r)=C_0\,\int \frac{\d r}{f\,R^2} +Const.\, ,
\end{equation}
where $C_0$ and $Const.$ are integration constants and without loss of generality we can set $Const.=0$. 

As we mentioned earlier first we want to find the simplest vacuum solution with scalar field (without having any other sources) which we already did in our previous researches \cite{Tahamtan-PRD-2015, Tahamtan2}, so we just summarize the results here 
\begin{eqnarray}
f(r)&=&1\, , \nonumber \\
R(r)&=&\sqrt{r^2-\chi^2}\, ,  \label{SF-solution}
\end{eqnarray}
and the scalar field is   
\begin{equation}\label{static-field}
\varphi(r)=\frac{C_{0}}{2\,\chi}\ln{\left\{\frac{r -\chi}{r+ \chi}\right \}}\, 
\end{equation}
and Einstein equations give condition $C_{0}=\sqrt{2}\chi$. Clearly, from (\ref{Ricci}) the solution has curvature singularity at $r=\chi$ which represents a naked time-like singularity. This is caused by diverging scalar field at this position. 

When $r \to \infty$ the scalar field is vanishing. The metric solution is also evidently asymptotically flat but the total area of spherical surfaces $r=const., t=const.$ grows quadratically with coordinate $r$ only far from the central region while close to the curvature singularity $r=\chi$ it grows just linearly.

 This solution coincides with certain parametric limit of Janis, Newmann and Winicour (JNW) solution \cite{JNW1} in the coordinates given in \cite{JNW2}. JNW solution is two-parametric and these parameters correspond to strength of scalar field and Schwarzschild-like mass.

\subsection{Nonlinear Electrodynamics}\label{NE-solution}
Here, we study one of the simplest models of NE called "Square Root" Lagrangian, namely $\mathcal{L}=-\sqrt{F}$. {Although this model does not possess Maxwell limit but, as already mentioned in Introduction, it has its own interesting properties. Particularly, this Lagrangian represents a strong field regime of many models of NE (e.g. Born--Infeld). } Note that in "Square Root" Lagrangian  when considering only the magnetic field all the energy conditions are satisfied unlike the case of pure radial electric field. 

Since our spacetime is static spherically symmetric, without loosing generality we assume the following electromagnetic field two-form
\begin{equation}
\textbf{F}=F_{\theta \phi}\, \d \theta \wedge \d \phi\, ,
\end{equation}
where $F_{\theta \phi}=q_{m}\sin{\theta}$ and $q_{m}$ can be considered as a magnetic charge. All the modified Maxwell equations are satisfied trivially. The electromagnetic invariant $F=F_{\mu \nu}F^{\mu \nu}$ becomes
\begin{equation}\label{F-NE}
F=\frac{2\,q^2_{m}}{R^4}\, .
\end{equation}
The energy momentum tensor (\ref{energy-momentum-Maxwell}) corresponding to $\mathcal{L}=-\sqrt{F}$ and the metric (\ref{ourmetric}) is
\begin{equation}\label{energy-momentum-NE}
{}^{\rm NE}T^{\mu}{}_{\nu}=diag\left\{-\frac{\sqrt{F}}{2},-\frac{\sqrt{F}}{2},0,0\right\}\, .
\end{equation}
For finding the solutions from Einstein equations ($G^{\mu}{}_{\nu}=T^{\mu}{}_{\nu}$), we first solve  $G^{t}{}_{t}=T^{t}{}_{t}$. Before that we can simplify the equations with respect to the symmetry in energy momentum tensor $T^{t}{}_{t}=T^{r}{}_{r}$ which leads to
\begin{equation}
  R_{,rr}=0
\end{equation}
{in which we can choose $R=r$. It has been shown in \cite{Jacobson2007} that the above condition on energy momentum tensor components leads to such result for any static spherically symmetric spacetime of four and more dimensions}. Now from (\ref{Gtt}) and (\ref{energy-momentum-NE}) we obtain
\begin{equation}
f+r\,f_{,r}-\alpha=0\, ,
\end{equation}
where $\alpha=1-\sqrt{2}\,q_m$, and the solution would be
\begin{equation}\label{f-NE}
f(r)=\alpha+\frac{C_{1}}{r}\, ,
\end{equation}
where $C_{1}$ is an integration constant and can be chosen as $-2m$, so $f(r)=\alpha-\frac{2m}{r}$. If $q_m=0$ then $\alpha=1$ and one would get Schwarzschild solution as should be expected in the absence of source. {Since this specific model of NE does not have Maxwell limit the obtained solution substantially differs from Reissner-Nordstr\"{o}m spacetime.}

Obviously for $\alpha$ positive, $f(r)$ has one zero root ($r_0$) which can be considered as an event horizon. One can check that at $r_0=\frac{2m}{\alpha}$ there is no true singularity. Kretschmann scalar is diverging only at the center, namely $r=0$.

 This solution is similar to the solution of geometry outside the core of so-called global monopole, a spacetime defect usually considered to be sourced by a self-coupling triplet of scalar fields whose original $O(3)$ symmetry is spontaneously broken to $U(1)$. Global monopole was discussed in detail and with many applications in literature, some of the original work can be found in \cite{Kibble, Letelier, Barriola, global1} and some recent work by author in \cite{Tahamtan, Boosting}.

Parameter $\alpha$ is related to solid angle deficit/excess which can be seen by the following transformation
\[\tilde{t}=\sqrt{\alpha}\,t,\quad \tilde{r}=\frac{r}{\sqrt{\alpha}},  \quad \tilde{m}= \frac{m}{{\alpha}^{3/2}}\, ,  \]
that gives the following line element
\begin{equation}\label{deficit}
\d s^2=-\left(1-\frac{2\tilde{m}}{\tilde{r}}\right)\,\d \tilde{t}^2+\frac{\d \tilde{r}^2}{\left(1-\frac{2\tilde{m}}{\tilde{r}}\right)} +\alpha\,\tilde{r}^2 \d {\Omega}^2\, .
\end{equation}
Asymptotically or for $\tilde{m}=0$ the relation between area of $\tilde{r}=const., \tilde{t}=const.$ surfaces and their proper radius clearly indicates deficit/excess of solid angle depending on the value of parameter $\alpha$. Obviously $\alpha$ should be positive but can be bigger or smaller than one depending on the sign of $q_m$. In the standard global monopole model with triplet of scalar fields only solid angle deficit is possible.

The above metric is not globally asymptotically flat. This NE model is interesting because it gives the geometry of global monopole in a much easier way and the solution is valid everywhere. 

\subsection{Einstein Maxwell Scalar Field }\label{EMS-solution}
For comparison with the solutions containing both NE and scalar field source that are derived in the following sections we present results for the Einstein--Maxwell system minimally coupled to a scalar field. As already mentioned in Introduction there are several studies regarding the Einstein--Maxwell scalar field solutions. Most of them used generating techniques to obtain solutions starting with scalar field spacetimes. For simplicity and easier comparison with forthcoming NE model (Square Root) we consider only purely magnetic field in static spherically symmetric spacetime.

The electromagnetic field (described by its only nonvanishing component) and electromagnetic invariant are the same as for general NE case, 
\[F_{\theta \phi}=q_{m}\sin{\theta},\quad\,\,\, F=\frac{2\,q^2_{m}}{R^4}\, .\]

The Maxwell energy momentum tensor corresponding to (\ref{energy-momentum-Maxwell}) and the metric (\ref{ourmetric}) is
\begin{equation}\label{energy-momentum-Max}
{}^{\rm Maxwell}T^{\mu}{}_{\nu}=\frac{F}{2}diag\left\{-1,-1,1,1\right\}\ .
\end{equation}
In this case, we proceed similarly to previous papers and generalize the static vacuum scalar field solution and assume
\begin{equation}\label{R-SMaxwell}
R(r)=\sqrt{\frac{(r-\chi_1)(r+\chi_2)}{f(r)}}\ ,
\end{equation}
with so far undetermined $f(r)$ and with two constants $\{\chi_1, \chi_2\}$. By using (\ref{box}) we arrive at the scalar field similar to the vacuum scalar field case (\ref{static-field}),
\[\varphi(r)=\frac{C_{0}}{\chi_1+\chi_2}\ln{\left\{\frac{r -\chi_1}{r+ \chi_2}\right \}}\, .\]
 Note that here $C_{0}$ and $\{\chi_1, \chi_2\}$ are independent constants unlike in (\ref{static-field}).

From $tt$ and $rr$ components of Einstein equations (using (\ref{field equations}) with (\ref{Scalar field}) and (\ref{energy-momentum-Max})), $G^{t}{}_{t}-G^{r}{}_{r}=T^{t}{}_{t}-T^{r}{}_{r}$, we can solve for the metric function $f(r)$ in the following form
\begin{equation} \label{f-SMaxwell}
f(r)=\frac{16\mu^2\,(\chi_1+\chi_2)^2\,\left(\frac{r-\chi_1}{r+\chi_2}\right)^{\mu}}{\left[\left(\frac{r-\chi_1}{r+\chi_2}\right)^{\mu}C_1-C_2\right]^2}\ ,
\end{equation}
where $\mu=\sqrt{1-\frac{C_0^2}{(\chi_1+\chi_2)^2}}$,
so $0<\mu\leq 1$. It is straightforward to check that the horizon at $r=\chi_{1}$ is singular (result known from previous studies of the solution). Without loosing generality one can consider $\chi_1=\chi_2=\chi$ and perform simple shift of radial coordinate which we assume further. The rest of Einstein equations are constraints for the integration constants, namely $C_1$ and $C_2$. They are bound to satisfy
\[C_1=\frac{4\,q_m^2}{C_2}\ .\]

This solution (\ref{f-SMaxwell}) corresponds to the one obtained in \cite{Penney1969} only written in different notation for the constants. Also, using simple relations for hyperbolic functions that are used to write the solution in \cite{Teixeira}, one can show that our solution corresponds to theirs. Note that in \cite{Teixeira}, they obtained the solution representing pure magnetic field using the duality between electric and magnetic field. As we will discuss subsequently, our solution has static vacuum solution limit. For solution discussed in \cite{JNW2} it was not possible to recover the static vacuum solution because one cannot make both electromagnetic and scalar fields vanish simultaneously --- the solution in \cite{JNW2} does not contain a branch enabling such limit.

It is clear that if $q_m$ vanishes we obtain static vacuum scalar field solution derived in \cite{JNW1} provided we demand $C_2=4\chi$ (needed in order to have correct Schwarzschild limit if we additionally remove scalar field by setting $C_{0}=0$).

If $C_0=0$ which means vanishing scalar field, we can introduce a new radial coordinate $\hat{r}=\frac{r(C_1-C_2)-\chi(C_1+C_2)}{4\,\chi}$ to recover Reissner-Nordstr\"{o}m solution as one would expect 
\[f(\hat{r})=1-\frac{2M}{\hat{r}}+\frac{q_m^2}{\hat{r}^2}\ .\]
Above, we have introduced a new parameter $M$ describing asymptotic mass with the following relation with the original constant $C_{2}$
\[\frac{4\,q_m^2}{C_2}+C_2=-4M\ .\]
By further removing electromagnetic field by setting $q_{m}=0$ we immediately recover Schwarzschild solution.

In both limiting procedures (first $q_{m}\to 0$ followed by $C_{0}\to 0$, or in reverse order) we arrive at the original vacuum solution --- the Schwarzschild black hole.

\subsection{Scalar Field and Square Root Lagrangian}\label{Exact-sol-SC-NE}
In this section we present an explicit solution for already considered NE model (Square Root) additionally minimally coupled to a massless scalar field. Note that the solution without scalar field obtained in section \ref{NE-solution} was a black hole with regular horizon so this model can serve as a test for the effects of scalar field on NE spacetimes (thereby probing the extended validity of ``Chase theorem'').

We start to solve the coupled system by considering $tt$ and $rr$ components of Einstein equations (\ref{field equations}), namely $G^{t}{}_{t}-G^{r}{}_{r}=T^{t}{}_{t}-T^{r}{}_{r}$. By inserting from (\ref{Gtt}), (\ref{Grr}), (\ref{Scalar field}) and (\ref{energy-momentum-NE}) to this equation we immediately obtain
\begin{equation}\label{SF-NE}
\varphi^2_{,r}=-\frac{2\,R_{,rr}}{R}.
\end{equation}
From the above equation and (\ref{phi}) we are able to find $f$ in terms of $R$ 
\begin{equation}\label{f}
f=\sqrt{-\frac{C^2_{0}}{2\,R^3\,R_{,rr}}}\ .
\end{equation}

The rest of Einstein equations will constrain the form of $R$. As we mentioned before, our model of NE is $\mathcal{L}=-\sqrt{F}$ with the energy momentum tensor (\ref{energy-momentum-NE}). From $G^{t}{}_{t}-({}^{\rm NE}T^{t}{}_{t}+{}^{\rm SF}T^{t}{}_{t})=0$, we get

\begin{equation}
f\,\left(\frac{R_{,r}}{R}\right)^2+\frac{R_{,r}}{R}\,f_{,r}-\frac{1}{R^2}+f\,\frac{R_{,rr}}{R}+\frac{q_m}{\sqrt{2}}\frac{1}{R^2}=0\, ,
\end{equation}
which together with (\ref{f}) gives the following expressions for $R$, $f$ and from (\ref{SF-NE}) for $\varphi$
\begin{eqnarray}
R(r)&=&\sqrt{\beta^2\,(r+\tilde{C_1})(r-\tilde{C_2})-C^2_0} \times \exp(-\Omega(r))\, ,\label{R-SC-NE}\nonumber\\
&&\\
f(r)&=&-\frac{e^{2\Omega(r)}}{\beta\,\sqrt{2}}\label{f-SC-NE}\, ,
\\
\varphi(r)&=&\frac{2\sqrt{2}C_0}{{\beta}(\tilde{C_1}+\tilde{C_2})}\,\Omega(r)\,,\label{SF-22}
\end{eqnarray}
where $\tilde{C_1}$ and $\tilde{C_2}$ are integration constants and we introduced parameters $\beta$, $\rho$ and a function $\Omega(r)$ in the following way
\begin{eqnarray}
\beta&=&(q_m-\sqrt{2})\, ,\\
\rho&=&\sqrt{\beta^2\,(\tilde{C_1}+\tilde{C_2})^2+4C^2_0}\, , \label{rho}\\
\Omega(r)&=&\frac{{\beta}(\tilde{C_1}+\tilde{C_2})}{\rho}\arccoth \left(\frac{\beta\,(2r+\tilde{C_1}-\tilde{C_2})}{\rho}\right) \nonumber\\
\end{eqnarray}
and $\beta$ should be negative for preserving the metric signature.

After presenting the explicit solution above, we will turn our attention to the investigation of existence of (event) horizon. Since the spacetime is static spherically symmetric one can check for zeros of the lapse function $f$. From (\ref{f-SC-NE}), it is sufficient that $\Omega \rightarrow -\infty$. Since $\beta$ is negative, the only way to achieve this is to assume $\tilde{C_1}+\tilde{C_2}\leq 0$ (we are constrained to negative branch of $\arccoth$). The argument of $\arccoth$ should then be $-1$ which happens for the following value of $r$

\[r_0=\frac{1}{2}\left(\tilde{C_2}-\tilde{C_1}-\rho/\beta \right)\, .\]

As is well-known, it is possible to write $\arccoth(x)$ in logarithmic form when $|x|>1$. Using that, first we write $\Omega$ in terms of $r_0$ as below
\begin{equation}
\Omega(r)=\frac{{\beta}(\tilde{C_1}+\tilde{C_2})}{2\,\rho}\ln \left(\frac{r-r_0}{r-\tilde{r}_0}\right)\, ,
\end{equation}
where $\tilde{r}_0=r_0+\rho/\beta$. After some simplifications, the equations (\ref{R-SC-NE}) and (\ref{f-SC-NE}) become
\begin{eqnarray}
R(r)&=&\sqrt{\beta^2\,(r-r_0)(r-\tilde{r}_0)} \left[\frac{r-\tilde{r}_0}{r-r_0}\right]^{\frac{\nu}{2}}\, ,\nonumber\\
 \label{R-SNE}&&\\
f(r)&=&-\frac{1}{\beta\,\sqrt{2}}\left[\frac{r-{r_0}}{r-\tilde{r}_0}\right]^{\nu}\, , \label{f-SNE}
\end{eqnarray}
where $\nu={\frac{|\beta(\tilde{C_1}+\tilde{C_2})|}{\rho}}\geq 0$. 

In this compact form it is clear that $f$ is vanishing at $r=r_0$ but the behavior of $R$ driven by the power of $(r-r_0)$ which is $\frac{\nu-1}{2}$. Depending on whether $\nu \lesseqqgtr 1$, $R$ would be zero, finite or diverge. Considering the definition for $\rho$ from (\ref{rho}), it is clear that $\nu<1$ if $C_{0}\neq 0$ (nontrivial scalar field), so for $r=r_0$ the function $R$ is vanishing.

Note that since $\beta$ is negative, $r_0>\tilde{r}_0$. So at $r_0$ there is an outermost horizon and it is potentially an event horizon but we need to see the behavior of Ricci scalar at $r=r_0$ to determine its regularity. From (\ref{Ricci})

\[Ricci \sim (r-r_0)^{\nu-2}\]
and since $\nu<1$ the Ricci Scalar at $r=r_0$ is clearly diverging. 

So in our solution, the event horizon is also a true singularity which confirms the role of scalar field in destroying horizon regularity even in this NE model. Thus our spacetime contains a null singularity along the horizon position which means that it is not possible to extend the spacetime and perform any analysis of SCC. On the other hand, since this is the stationary state of geometry it shows that the no-hair theorem is valid in this case as well since we have not found black hole spacetime with both nongravitational fields being nontrivial.

The scalar field (\ref{SF-22}) becomes 
\begin{equation}\label{SF-NE-model}
\varphi(r)=\frac{\sqrt{2}C_0}{\rho}\ln \left[\frac{r-{r_0}}{r-\tilde{ r}_0}\right]
\end{equation}
and it is clear that at $r=r_0$, the scalar field is diverging as well and the same applies to electromagnetic invariant (\ref{F-NE}) and therefore to NE energy momentum tensor (\ref{energy-momentum-NE}). 

The obtained solution, (\ref{R-SNE}) and (\ref{f-SNE}), is a NE generalization of Janis, Newmann and Winicour solution \cite{JNW1} and the original solution is recovered for $q_m=0$ {while as well setting  $\tilde{C_1}=\tilde{C_2}$}.

{If we consider that scalar field vanishes, $C_0=0$, then necessarily $\nu=1$ and the solution in (\ref{R-SNE}) and (\ref{f-SNE}) will be equivalent to \ref{f-NE} upon trivial changes in coordinates and constants --- such as $R \rightarrow \beta (r+\tilde{C_1})$, introducing $\alpha=-\beta \sqrt{2}$ and redefining additional constants to obtain proper mass parameter).  }

If we assume $\tilde{C_1}$ and $\tilde{C_2}$ are zero then the form of the metric functions is simpler 
\begin{eqnarray}
R(r)&=&\sqrt{\beta^2\,r^2-C^2_0}\, , \\
f(r)&=&-\frac{1}{\beta\,\sqrt{2}}\, ,
\end{eqnarray}
leading to time-like naked singularity. When $q_m$ in $\beta$ vanishes then the solution becomes equivalent to (\ref{SF-solution}) with some trivial redefinition of coordinate $r$.

 Both the general metric solution and its subcases with nontrivial scalar field  do not have any regular horizon. {Although there are clear differences between Maxwell and Square Root Lagrangian solutions with scalar field they both contain singular horizon or naked singularity due to the scalar field presence. Direct comparison of Square root model with Maxwell case is difficult since this NE model does not have Maxwell weak field limit or similar strong field behavior. However, from the form of function $R(r)$ in both cases ((\ref{R-SNE}) and \ref{R-SMaxwell}) we see that the behavior of areal radii is similar (the same applies to the form of scalar fields) while there is substantial difference in the form of metric function $f(r)$ (see (\ref{f-SNE}) and (\ref{f-SMaxwell})). Note that there is profound difference already for solutions without scalar field as Square root model metric (\ref{deficit}) only possesses single horizon compared to Reissner--Nordstr\"{o}m solution with two and global asymptotics varies as well. Nevertheless, the scalar field produces singular horizons in both cases.}

\section{ No Black hole solutions}\label{General-source}
For highly symmetric static spacetime to represent a black hole its lapse function should have at least one root. But this root should not be a true singularity at the same time. In other words, at least the Ricci scalar and Kretschmann scalar at this "root" should not diverge. Here we consider a presence of horizon for larger class of sources and the influence scalar field exerts in such situation in order to probe the ``Chase theorem'' extension.

For this purpose, we consider as before the radial scalar field and additionally any other kind of source with the condition ${}^{\rm Other}T^{t}{}_{t}={}^{\rm Other}T^{r}{}_{r}$ (e.g., satisfied by NE). { If ${}^{\rm Other}T^{r\mu}_{\nu}$ was the only source, this condition would automatically imply $g_{tt}g_{rr}=-1$ according to \cite{Jacobson2007}. However, here the total energy momentum tensor does not satisfy such constraint.} Before starting the analysis we choose a new coordinate system which makes the subsequent derivation easier. Therefore we start with the following metric 
\begin{eqnarray}\label{metric}
\d s^2&=&-f(r)\,\d t^2+\frac{h(r)}{f(r)}\, \d r^2+r^2 \d {\Omega}^2\, .
\end{eqnarray}
The metric density is
\[\sqrt{-g}=\sqrt{h}\,r^2\sin \theta\, ,\]
which constrains the metric function $h$ to be positive. 

By following the same procedure as before, using scalar field wave equation (\ref{box}) and  $G^{t}{}_{t}-G^{r}{}_{r}=T^{t}{}_{t}-T^{r}{}_{r}$ field equations of Einstein equations (\ref{field equations}), we find the following expression for the scalar field and the lapse function,
\begin{eqnarray}\label{}
\varphi(r)&=&\int \sqrt{\frac{h_{,r}}{rh}}\,\d r\, , \\
f&=&\frac{f_0\,h}{\sqrt{r^3\,h_{,r}}}\, ,
\end{eqnarray}
where $f_0$ is an integration constant. 
Ricci Scalar for this metric (\ref{metric}) is
\begin{eqnarray}\label{Ricci-sc}
&&Ricci=\frac{2}{r^2}+\frac{f_0}{4\,\sqrt{r^5\,h_{,r}}}\left\{\boxed{2\,r\,{\left(\frac{h_{,r}}{h}\right)}^2+\frac{h_{,r}-r\,h_{,rr}}{h}} \right.\nonumber \\
&&+\left.\frac{1}{r\,(h_{,r})^2}\left[r^2\left(2{(h_{,rrr})}\,h_{,r}-3{(h_{,rr})}^2\right)+(r\,{h_{,r}}^2)_{,r}\right]\right\}\nonumber .\\
\end{eqnarray}
When the lapse function $f$ is zero (which occurs when $h=0$), there is a possibility to have horizon(s). So we are assuming that at $r=r_0$ there is at least one zero for $h$ and the first derivative of $h$ with respect to $r$ at this point is non zero and finite. As one can see from Ricci scalar (\ref{Ricci-sc}) vanishing $h$ means diverging curvature. Even if we assume the expression in the box containing $h$ in denominator to vanish identically it leads to more problems because from  
\[2\,r\,{\left(\frac{h_{,r}}{h}\right)}^2+\frac{h_{,r}-r\,h_{,rr}}{h}=0\]
we obtain the following solution for $h$
\[h=-\frac{h_0}{r^2+h_1}\, ,\]
where $h_0$ and $h_1$ are integration constants. Ruling out purely imaginary solution for $f$, the constant $h_0$ must be positive which makes the metric density imaginary. Even without caring about reality conditions this form of $h$ makes the solution automatically a naked singularity one since it leads to $f\,{\sim}\, 1/r^2$. 

All these discussions show that in the presence of minimally coupled massless scalar field in addition to the other type of source satisfying ${}^{\rm Other}T^{t}{}_{t}={}^{\rm Other}T^{r}{}_{r}$ (potentially even those forms of NE that remove curvature singularities thus creating so-called regular black holes) it is impossible to have black hole solutions with regular horizon.

\section{Test Field}\label{Test-field}

In this section we use a test scalar field approach to learn more about the scalar fields behavior in the vicinity of horizon. First, inspired by \cite{Herdeiro} where the comparison between test electric field (in Maxwell theory) and a test scalar field on a black hole spacetime was performed, we apply this reasoning to NE theory instead of source-free Maxwell theory. Subsequently, we investigate the scalar field wave equation on a generic black hole spacetime taking into account complete set of solutions, not restricting ourselves to purely spherically symmetric ones.

\subsection{Difference between Electric and Scalar Field as a test fields}
 
 Since Klein--Gordon equation and Maxwell (also modified Maxwell) equations for potential are similar it is worth to see what is the difference between them which leads to the result that the Maxwell field obeys a Gauss law and scalar field does not. 
 
 For showing the difference between the electric and scalar fields on fixed background (\ref{ourmetric}) (assuming $R=r$ {which is always possible when $T^{t}{}_{t}=T^{r}{}_{r}$ according to \cite{Jacobson2007}}), we consider a test, spherically symmetric electromagnetic field described by the gauge potential one-form
\begin{equation}
\textbf{A}=\phi(r)\, \d t ,
\end{equation}
then $\textbf{F}=-\phi_{,r}\, \d t \wedge \d r$. From (\ref{modified Maxwell}), we get 

\begin{equation} \label{com-NE}
\partial _{r}(r^2\sin{\theta}\mathcal{L}_F\,\phi_{,r})=0\,\rightarrow \, \phi_{,r}=\frac{Q_{E}}{r^2\mathcal{L}_F}\, .
\end{equation} 
Note that in Maxwell theory $\mathcal{L}_F=-1$. One can write the above equation in the following form 
\begin{equation}\label{compare-NE} 
\partial _{r}(r^2\sin{\theta}\,\phi_{,r})=-r^2\sin{\theta}\,\phi_{,r}\,(\ln{\lvert\mathcal{L}_F\rvert})_{,r}\, ,
\end{equation}
which looks like ordinary Maxwell equation with source.

In the case of radial scalar field, Klein-Gordon equation (\ref{box}) simplifies into
\begin{equation} \label{com-SF}
\partial _{r}(r^2\sin{\theta}f(r)\,\varphi_{,r})=0\,\rightarrow \,\varphi_{,r}=\frac{Q_{SF}}{r^2\,f(r)}\, ,
\end{equation}
with $Q_{SF}$ being integration constant with the interpretation of scalar charge. It is possible to write the above equation (\ref{com-SF}) in similar way as the NE equation (\ref{compare-NE}) in source form, that is
\begin{equation} \label{compare-SF}
\partial _{r}(r^2\sin{\theta}\,\varphi_{,r})=-r^2\sin{\theta}\,\varphi_{,r}\,(\ln \lvert f\rvert)_{,r}\, .
\end{equation}

From (\ref{com-SF}) it is clear that at the event horizon which corresponds to $f(r_{0})=0$, the scalar field is diverging while in case of electromagnetic field (\ref{com-NE}) the fields are regular on the horizon both in the Maxwell case and for generic NE model case. The different behavior of these two fields is due to the presence of lapse function in scalar field equation in contrast to electromagnetic field equation. Note that in denominator of (\ref{com-NE}) there is $\mathcal{L}_F$ (which is some function in $F$). In general, NE models are used to remove any kind of singularity in "Electric" field and as a result in $F$ which in turn means $\mathcal{L}_F$ on the horizon should not vanish. So considering a special model of NE which would provide vanishing $\mathcal{L}_F$ on the horizon and thus behave effectively similarly to scalar field (compare (\ref{compare-NE}) and (\ref{compare-SF})) goes counter to the standard consideration in NE models and cannot be reasonably justified. In the case of scalar field on the other hand this behavior is pretty robust and not dependent on specific fine-tuning of model \cite{Tafel}. Still one can ask if such a specific NE model with divergence on the horizon might counter the scalar field divergence. However, looking at their corresponding energy momentum tensors (\ref{energy-momentum scalar-field}, \ref{energy-momentum-Maxwell}) and considering that $\mathcal{L}_F\rvert_{r=r_{0}}=0$ at the same time one can see that it is not possible to cancel the role of scalar field in causing horizon irregularity even by the fine-tuning when the backreaction on geometry is considered.

\subsection{Wave equation for Black hole solutions close to event horizon}

We have studied "Chase Theorem" in exact solutions with other sources in parts \ref{Exact-sol-SC-NE} and \ref{General-source} with certain limitations (spherically symmetric scalar field specific form of NE or with additional sources satisfying certain symmetry of energy momentum tensor). Now, we will study test scalar field without the assumption of symmetry for completely generic case on the background of SSS geometry to confirm our previous results. This study is in line with the analysis performed already by Chase in \cite{Chase} for Schwarzschild background.

We consider a scalar perturbation obeying the Klein-Gordon equation $\Box \Psi(t,r,\theta,\phi)=0$ for the static spherically symmetric geometry (\ref{ourmetric}) with $R=r$ {(always possible for sources satisfying $T^{t}{}_{t}=T^{r}{}_{r}$ \cite{Jacobson2007})}
\begin{eqnarray}\label{wave}
f\,\Psi_{,rr}+\left(f_{,r}+\frac{2f}{r}\right)\Psi_{,r}-\frac{\Psi_{,tt}}{f}\, ,\nonumber\\
+\frac{1}{r^2}\left\{\Psi_{,\theta \theta}+\cot{\theta}\Psi_{,\theta}+\frac{\Psi_{,\phi \phi}}{\sin^2{\theta}}\right\}=0\, .
\end{eqnarray}
For the above equation one  can use the symmetry of background and employ usual separation of variables 
\begin{equation}
\Psi(t,r,\theta,\phi)=e^{\pm \textrm{i}\omega t}\,\frac{\psi(r)}{r}\,Y^m_{l}(\theta,\phi)\, ,
\end{equation}
where $Y^m_{l}$ is the harmonic function on the unit 2-sphere. This kind of separation of variables (identical to the one in \cite{Chase}) is selected since it allows to study static situation as well. On the other hand, in the studies of quasi-normal modes a decomposition based on null coordinates is preferable since the boundary conditions at horizon and infinity are straightforward to impose.

With this assumption we would have the radial equation as follows
\begin{equation}\label{Radial}
f\,\psi_{,rr}+f_{,r}\,\psi_{,r}-\left(\frac{l(l+1)}{r^2}-\frac{\omega^2}{f}+\frac{f_{,r}}{r}\right)\psi=0\, .
\end{equation}
Since we are interested in perturbation around non-degenerate horizon, we assume 
\[f=f_0\,(r-r_0)+O((r-r_0)^2)\, ,\]
where $f_0$ is a positive constant (e.g., $f_{0}=\frac{1}{2m}$ in Schwarzschild). Upon substituting into (\ref{Radial}) we obtain 
\begin{eqnarray}
  f_0(r-r_0)\,\psi_{,rr}+f_0\,\psi_{,r}-\qquad\qquad\qquad\qquad\\
 \qquad -\left(\frac{l(l+1)}{r^2}-\frac{\omega^2}{f_0(r-r_0)}+\frac{f_0}{r}\right)\psi=0\, .\nonumber
\end{eqnarray}
The solution for the above equation is of the form
\begin{eqnarray}\label{wave-final}
\psi&=&(r-r_0)^{\frac{-\textrm{i}\,\omega}{f_0}}\left[\psi_0\,r^{n_{1}}{}_{2}F_{1}\left(a_1,b_1;n_1;\frac{r}{r_0}\right) \right. \nonumber\\
&&\left.+\psi_1\,r^{n_2}{}_{2}F_{1}\left(a_2,b_2;n_2;\frac{r}{r_0}\right)\right]\, ,
\end{eqnarray}
with ${}_{2}F_{1}$ being hypergeometric function. And $a_i, b_i, n_i$ where $i=1,2$ are the parameters defined as following
\begin{eqnarray}
a_i&=&\frac{\left(1-\sigma_i \right)}{2} -\frac{\textrm{i}\,\omega-\sqrt{f_0^2-\omega^2}}{f_0}\, ,\nonumber\\
b_i&=&\frac{\left(1-\sigma_i \right)}{2} -\frac{\textrm{i}\,\omega+\sqrt{f_0^2-\omega^2}}{f_0}\, ,\nonumber\\
n_i&=&1-\sigma_{i}\, ,\label{F-parameters}\\
\sigma_i&=&(-1)^{i+1}\,\sqrt{1-\frac{4\,l(l+1)}{f_0\,r_0}}\, ,\nonumber
\end{eqnarray}
these parameters clearly satisfy the following relation
\begin{equation}\label{param-relation}
  n_i-(a_i+b_i)=\frac{2\,\textrm{i}\,\omega}{f_0}\, .
\end{equation}
The solution contains a factor $(r-r_0)^{\frac{-\textrm{i}\,\omega}{f_0}}$ leading to divergence of the derivative of the following form $\psi_{,r}\sim (r-r_0)^{-1}$ because the asymptotic behavior of hypergeometric function around $r_0$ (or in other words around $\frac{r}{r_0}=1$) \cite{wolf} has the following form
\begin{eqnarray}
&&{}_{2}F_{1}\left(a_i,b_i;n_i;\frac{r}{r_0}\right) \propto \\ &&D_0\,(r_0-{r})^{\frac{2\,\textrm{i}\,\omega}{f_0}}\left\{1+D_1\,({r}-{r_0})+O(r-r_0)^2\right\}+ \nonumber\\
&&D_2\left\{1+D_3\,(r-r_0)+O(r-r_0)^2\right\}\, .\nonumber
\end{eqnarray} 
All these $D_j$ are constant in terms of $a_i, b_i, n_i$.
This in turn leads to divergence in the stress energy momentum tensor of the scalar field (\ref{Scalar field}) which shows that generic test scalar field blows up on the horizon. One can take this as a confirmation of the Chase theorem or as an indication of the need to study this problem including the backreaction of the scalar field on the geometry as we did in previous parts.

The above analysis seemingly fails for $\omega=0$. However, if we consider $\omega=0$ in the formula (\ref{param-relation}) for relation between parameters (\ref{F-parameters}) we get zero on the right-hand side and moreover all parameters become real. Since we are looking for the limit when we get close to horizon $r_0$ we will check the asymptotic behavior of hypergeometric function close to one which attains modified form due to the simplification in parameters  
\begin{eqnarray}
&&{}_{2}F_{1}\left(a_i,b_i;(a_i+b_i);\frac{r}{r_0}\right) \propto \nonumber\\ 
&&\tilde{D_0}\left(\log{(1-\frac{r}{r_0})}+\tilde{D_1}\right)\left\{1+O(r-r_0)\right\}\, .\nonumber
\end{eqnarray}
This again leads to divergence in scalar field which is moreover consistent with the logarithmic behavior in (\ref{SF-NE-model}) which should be the case for static mode.

\section{Conclusion and final remarks}
The role of scalar field is so dominant that it can effect the spacetime solution drastically by turning it from black hole solution into naked singularity or causing the horizon to become singular. We have studied this process in the presence of additional fields (especially NE field) in several ways. {In order to have comparison between Square Root model and Maxwell theory, first we revisited Einstein--Maxwell scalar field solution.} In case of NE model with Lagrangian $\sim \sqrt{F}$ which captures the strong field regime of many NE models (e.g., Born--Infeld) we have derived solution which generalizes that of Janis--Newman--Winicour and shows irregular horizon. {Although, the solution differs substantially from the Maxwell scalar field solution the regularity of horizon is not improved.} For vanishing scalar field the solution represents geometry of so-called global monopole which in this case can have both deficit and excess solid angle. 

Subsequently, we have shown that regular horizon absence features in much larger class of additional sources when scalar field is present. Finally, we have studied the horizon regularity in the test field approximation which brings better understanding of the scalar field influence and enables treatment of scalar fields that are not spherically symmetric. The test scalar field approach confirmed the previous results.

All the results are in agreement with Chase theorem and show that there are no static black holes with regular horizon in these models. This shows that no-hair theorem should be valid in such a scenarios.

The exact solutions with scalar field studied here posses either null singularity or in special cases a time-like one. This means that they either violate Weak Cosmic Censorship or are infinitely close to it. Since in generic cases with NE one can not pass the exterior horizon because it is singular it is not possible to study processes close to potential Cauchy horizon and thereby Strong Cosmic Censorship. 

In future, we will study massive scalar field with potentials to understand if the behavior is as generic as in the case of scalar-vacuum studied in \cite{Tafel}. For understanding SCC one should study dynamical solutions with electromagnetic and scalar fields.

\section*{Acknowledgments}
We would like to thank Vitor Cardoso and Otakar Sv\'{\i}tek  for enlightening discussions and useful suggestions. We are as well grateful to referee for numerous improvements of this manuscript. This work was supported by the research grant GA\v{C}R 17-16287S, the INTER-EXCELLENCE project No.LTI17018 that supports the collaboration between the Silesian University in Opava and the Astronomical Institute in Prague.


\begin{thebibliography}{25}
	
\bibitem{penros-65} R. Penrose, {\em Phys. Rev. Lett.} \textbf{14}, 57–59 (1965).

\bibitem{penros-69} R. Penrose, {\em Riv. Nuovo Cim. }\textbf{1}, 252–276 (1969). [{\em Gen. Rel. Grav.} \textbf{34},1141(2002)].
	
\bibitem{Hawking} S. W. Hawking and R. Penrose, {\em Proc. Roy. Soc. Lond.} \textbf{A314}, 529–548 (1970). 

\bibitem{Cardoso1} V. Cardoso, J. L. Costa, K. Destounis, P. Hintz, and A. Jansen, {\em Phys. Rev. Lett.} \textbf{120}(3), 031103 (2018).

\bibitem{Cardoso2} R. Luna, M. Zilhao, V. Cardoso, J. L. Costa, and J. Natario, {\em Phys. Rev. D} \textbf{99} (6), 064014 (2019). 

\bibitem{Hod} S. Hod, {\em Nucl. Phys.} \textbf{B941}, 636–645 (2019). 
	
\bibitem{Chase} J. E. CHASE, {\em Commun. math. Phys.} \textbf{19}, 276-288 (1970).

\bibitem{Tafel} J. Tafel, {\em Gen. Rel. Grav.} \textbf{46}, 1645 (2014).

\bibitem{Herdeiro} C. A. R. Herdeiro and Eugen Radu,{\em Int. J. Mod. Phys. D} \textbf{24}, 1542014 (2015).

\bibitem{Dirac} P.A.M. Dirac, {\em Lectures on Quantum Mechanics}, (Yeshiva University, New York 1964).

\bibitem{BornInfeld} M. Born, L. Infeld, {\em Proc. R. Soc. (London)} A {\bf 144}, 425 (1934).

\bibitem{Plebanski} J. Pleba\'{n}ski, {\em Lectures on Non-linear Electrodynamics}, (Nordita 1970).

\bibitem{Ayon-Beato1} E. Ay\'{o}n-Beato, A. Garc\'{\i}a, {\em Phys. Rev. Lett. }{\bf 80}, 5056 (1998).

\bibitem{Ayon-Beato2} E. Ay\'{o}n-Beato, A. Garc\'{\i}a, {\em Phys. Lett. B} {\bf 464}, 25 (1999).

\bibitem{Bronnikov} K. A. Bronnikov, {\em Phys. Rev. D} {\bf 63}, 044005 (2001).

\bibitem{china} Z. Y. Fan and X. Wang, {\em Phys. Rev. D} {\bf 94}, 124027 (2016).

\bibitem{SCC-Born} Qingyu Gan, Guangzhou Guo, Peng Wang and Houwen Wu, {\em Phys. Rev. D} {\bf 100}, 124009 (2019).

\bibitem{JNW1} A.I. Janis, E.T. Newman, and J. Winicour, {\em Phys. Rev. Lett.} \textbf{20}, 878 (1968);
\\M. Wyman,  {\em Phys. Rev. D} {\bf 24}, 839 (1981);
\\ H. A. Buchdahl, {\em Phys. Rev.} \textbf{111}, 1417 (1959);
\\ O. Bergman and R. Leipnik, {\em Phys. Rev.} \textbf{107}, 1157 (1957).

\bibitem{Fisher} I.Z. Fisher, {\em Scalar mesostatic field with regard for gravitational effects}, {\em Zh. Eksp. Teor. Fiz.} \textbf{18}, 636 (1948), English translation: gr-qc/9911008.

\bibitem{Penney1969} R. Penney, {\em Phys. Rev.} \textbf{182}, 1383 (1969).

 \bibitem{Uhlir1973} M. Uhl\'{\i}\v{r} and J. Dittrich, {\em Czech. J. Phys. B} {\bf 23}, 1 (1973).
 
\bibitem{Teixeira} A. F. da F. Teixeira, Idel Wolk, and M. M. Som, {\em J. Math. Phys.} {\bf 15}, 1756 (1974); {\em J. Phys. A: Math. Gen.} {\bf 9}, 53 (1976).

\bibitem{Gurses1977} A. Eris and M. G\"{u}rses, {\em J. Math. Phys.} {\bf 18}, 1303 (1977);
\\A. Banerjee and S. B. Dutta Choudhury, {\em Phys. Rev. D} {\bf 15}, 3062, (1977).

\bibitem{Maeda2019} H. Maeda and C. C. Mart\'{i}nez, {\em Class. Quantum Grav.} {\bf 36}, 185017 (2019).

\bibitem{JNW2} A.I. Janis, D.C. Robinson, and J. Winicour, {\em Phys. Rev.} \textbf{186}, 1729 (1969).

\bibitem{Roy}A. R. Roy and C. R. Datta,{\em Commun. math. Phys.} {\bf 29}, 285 (1973);
\\ N. Van den Bergh, {\em Gen. Relativ. Gravit.} {\bf 15}, 449 (1983).

\bibitem{Roberts} M.D. Roberts, Gen. Rel. Grav. \textbf{21}, 907 (1989).

\bibitem{Tahamtan-PRD-2015} T. Tahamtan and O. Sv\'{\i}tek, {\em Phys. Rev. D} {\bf 91}, 104032 (2015).

\bibitem{Tahamtan2} T. Tahamtan, O. Sv\'{\i}tek, {\em Phys. Rev. D} {\bf 94}, 064031 (2016).

\bibitem{Nielsen1973} H.B. Nielsen and P. Olesen, {\em Nucl. Phys. B} {\bf 57}, 367 (1973).

\bibitem{Hooft} G. ‘t Hooft, {\em Nucl. Phys. B} {\bf 121}, 333 (2003).

\bibitem{Guendelman2006}P. Gaete and  E. Guendelman, {\em Phys. Lett. B} {\bf 640}, 201 (2006).

\bibitem{Guendelman2014} M. Vasihoun and  E. Guendelman, {\em Int. J. Mod. Phys. A} {\bf 29}, 1430042 (2014).

\bibitem{square root} H. Maeda, M. Hassa\"{i}ne, C. Mart\'{i}nez, {\em Phys. Rev. D} {\bf 79}, 044012 (2009);\\
S.H. Hendi,  {\em Phys. Rev. D} {\bf 82}, 064040 (2010),\\
O. Gurtug, S.H.  Mazharimousavi, M. Halilsoy, {\em Phys. Rev. D} {\bf 85}, 104004 (2012).

\bibitem{Jacobson2007} T. Jacobson, {\em Class. Quant. Grav.} {\bf 24}, 5717 (2007).

\bibitem{Kibble} T. W. B. Kibble, {\em Phys. A} \textbf{9}, 1387 (1976). 
\bibitem{Letelier} P.S. Letelier, {\em Phys. Rev. D} \textbf{20}, 1294 (1979).
\bibitem{Barriola} M. Barriola, A. Vilenkin, {\em Phys. Rev. Lett.} \textbf{63}, 341 (1989).
\bibitem{global1} D. Harari and C. Lousto, {\em Phys. Rev. D} {\bf 42} 2626(1990).

\bibitem{Tahamtan} T. Tahamtan, O. Sv\'{\i}tek, {\em Eur. Phys. J. C} \textbf{74}, 2987 (2014).

\bibitem{Boosting} O. Sv\'{\i}tek and T. Tahamtan, {\em Gen. Rel. Grav.} \textbf{48}, 22 (2016).

\bibitem{wolf} http://functions.wolfram.com/HypergeometricFunctions\\/Hypergeometric2F1/02/01/



\end{thebibliography}
\end{document}